\documentclass[]{AO4ELT}  %>>> use for US letter paper
\usepackage{microtype}
\usepackage{biblatex}
\usepackage{amsmath,amsfonts,amssymb}
\usepackage{graphicx}
\usepackage{pst-all} % Pour pstricks
\usepackage[colorlinks=true, allcolors=blue]{hyperref}
\makeatletter         
\def\@maketitle{
\includegraphics[width = 170mm]{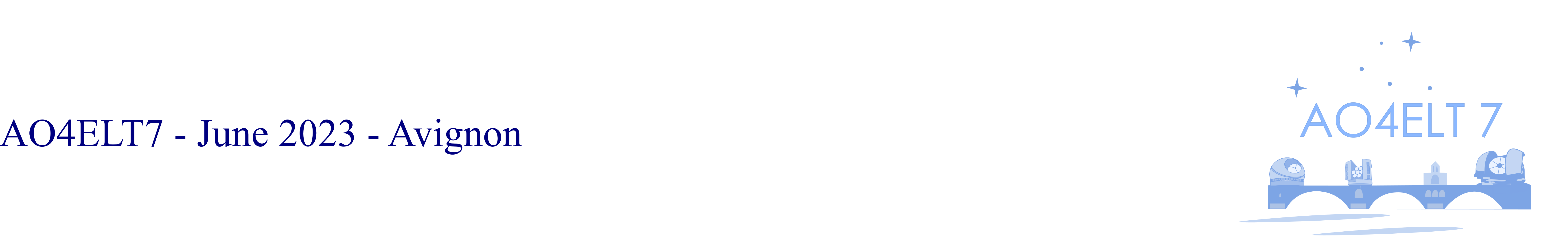}\\[8ex]
\begin{center}
{\Huge \bfseries \sffamily \@title }\\[4ex] 
{\Large  \@author}\\[4ex] 
\@date%\\[8ex]
\end{center}}
\title{Spectroastrometry and Imaging Science with Photonic Lanterns on Extremely Large Telescopes}

\author[a]{Yoo Jung Kim}
\author[a]{Michael P. Fitzgerald}
\author[a]{Jonathan Lin}
\author[b]{Steph Sallum}
\author[c]{Yinzi Xin}
\author[c]{Nemanja Jovanovic}
\author[d]{Sergio Leon-Saval}
\author[d]{Christopher Betters}
\author[c]{Pradip Gatkine}
\author[e]{Olivier Guyon}
\author[e]{Julien Lozi}
\author[c]{Dimitri Mawet}
\author[d]{Barnaby Norris}
\author[e]{Sebastien Vievard}

\affil[a]{Department of Physics and Astronomy, University of California, Los Angeles, 475 Portola Plaza, Los Angeles, CA 90095, USA}
\affil[b]{Department of Physics and Astronomy, University of California, Irvine, 4129 Frederick Reines Hall, Irvine, CA 92697, USA}
\affil[c]{Department of Astronomy, California Institute of Technology, 1200 East California Boulevard, Pasadena, CA 91125, USA}
\affil[d]{School of Physics, University of Sydney, Camperdown NSW 2006, Australia}
\affil[e]{Subaru Telescope, National Observatory of Japan, HI 96720, USA}

\authorinfo{Further author information: (Send correspondence to Y.J.Kim)\\Y.J.Kim.: E-mail: yjkim@astro.ucla.edu}

\begin{document} 
\maketitle
\begin{abstract}
Photonic lanterns (PLs) are tapered waveguides that gradually transition from a multi-mode fiber geometry to a bundle of single-mode fibers. In astronomical applications, PLs can efficiently couple multi-mode telescope light into a multi-mode fiber entrance and convert it into multiple single-mode beams. The output beams are highly stable and suitable for feeding into high-resolution spectrographs or photonic chip beam combiners. For instance, by using relative intensities in the output cores as a function of wavelength, PLs can enable spectroastrometry. In addition, by interfering beams in the output cores with a beam combiner in the backend, PLs can be used for high-throughput interferometric imaging. When used on an Extremely Large Telescope (ELT), with its increased sensitivity and angular resolution, the imaging and spectroastrometric capabilities of PLs will be extended to higher contrast and smaller angular scales. We study the potential spectroastrometry and imaging science cases of PLs on ELTs, including study of exomoons, broad-line regions of quasars, and inner circumstellar disks.
\end{abstract}


\section{INTRODUCTION}
\label{sec:intro} 

Achieving a smaller angular resolution is one of the key goals in astronomy, which enables resolving smaller objects at greater distances and the detection of companions at smaller separations. Even with 30\,m-class extremely large telescopes (ELTs), there are many applications in astronomy that will benefit from the ability to achieve sub-diffraction limit angular resolution, such as detecting exomoons, resolving broad-line region of quasars, imaging circumstellar disks at optical and near-infrared wavelengths, and detecting accreting planets.

\begin{figure} [ht]
\begin{center}
\begin{tabular}{c} 
\includegraphics[width=1\linewidth]{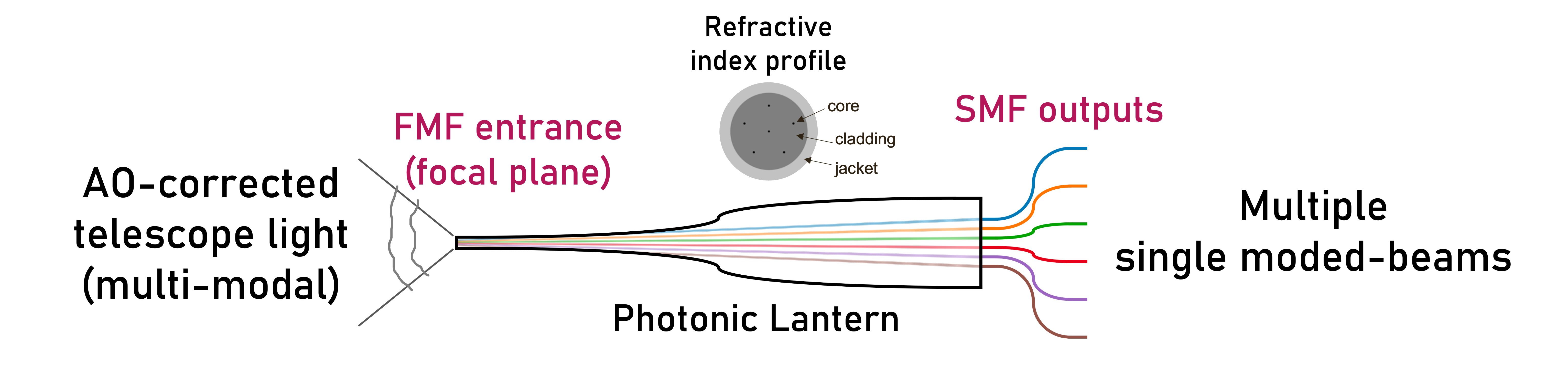}
\end{tabular}
\end{center}
\caption[example] 
{ \label{fig:PL} 
A schematic of a 6-port photonic lantern. AO-corrected telescope light is coupled into the lantern FMF entrance at the focal plane, gets confined into the cores, and the outputs are multiple single-moded beams on the other end.}
\end{figure}

The Photonic lantern (PL) is a photonic device that may be used for detecting signals and imaging structures smaller than the diffraction limit. Figure \ref{fig:PL} shows a schematic diagram of a PL. PLs are tapered waveguides that gradually transition from a few-mode fiber (FMF) geometry to a bundle of single-mode fibers (SMFs), or a few-core fiber (FCF) \cite{leo13, bir15}. The FMF entrance of the PL is placed at the telescope focal plane where AO-corrected light is coupled. As the light propagates through the lantern transition, it gets confined within the cores with low loss. Thus, a PL splits a few-moded wavefront into a few single-moded beams, each in an individual SMF. 

\begin{figure} [ht]
\begin{center}
\begin{tabular}{c} 
\includegraphics[width=0.7\linewidth]{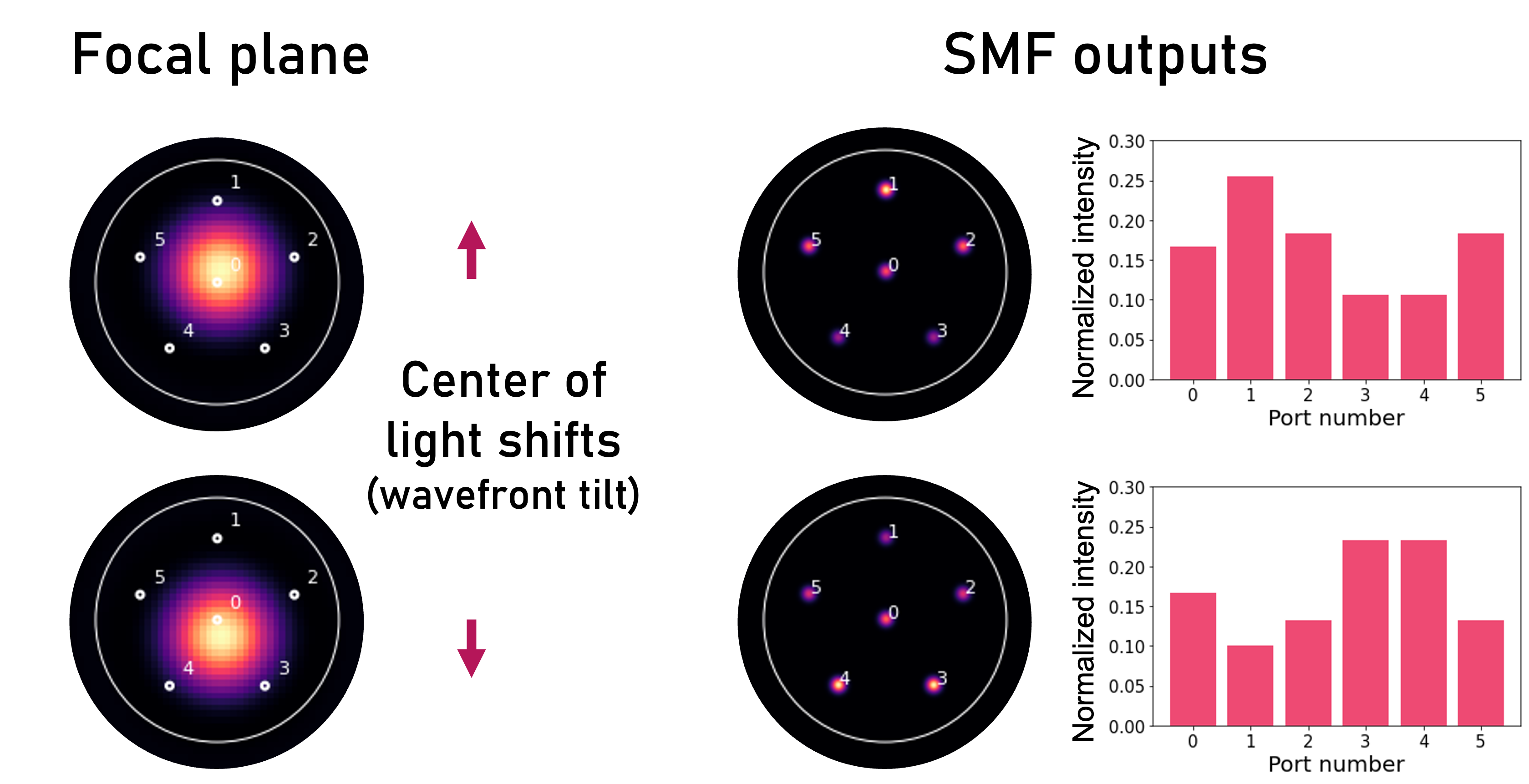}
\end{tabular}
\end{center}
\caption[example] 
{ \label{fig:demo} 
(Left) Light at the focal plane (lantern entrance) with centroid shift and (right) corresponding SMF outputs (lantern output). The relative intensities in the SMF outputs vary with the position of the light centroid. Thus the normalized intensities can be used to infer the source intensity distribution.}
\end{figure}

PLs do not split fluxes equally to each SMF output; rather, each output's complex amplitude corresponds to the wavefront filtered by some unique complex-valued aperture, determined by the pupil plane lantern principal modes (Kim et al., submitted \cite{kim23}). Thus the complex amplitudes in SMF outputs encode information on the input wavefront.
% PLs do not split fluxes equally to each SMF output but the relative fluxes in the SMFs depend on the input wavefront.
For instance, Figure \ref{fig:demo} shows the relative intensities in output SMFs for two input scenes: small tilts in the input wavefront. 
% Each PL output corresponds to some unique complex-valued aperture \cite{kim23}, so the PL outputs encode information on the input wavefront.
Using this feature, the application of PLs as a focal-plane wavefront sensor has been studied and demonstrated \cite{lin22a, nor20}. The single-moded PL outputs can be fed into a high-resolution spectrometer \cite{jov16, lin21b}. Spectroastrometric methods can be used with PLs by sensing wavelength-dependent relative intensities in the output SMFs \cite{kim22, lev23}. By interfering the SMF outputs in a backend integrated circuit beam combiner and measuring interferometric observables, smaller angular resolution may be achieved (Kim et al., submitted \cite{kim23}). 

In this paper, we explore the potential science applications of a PL for high angular resolution astronomy using two techniques: spectroastrometry and interferometric imaging.

% \begin{itemize}
% \item intro of PL
% \item sensitivity to small angular scales (the First figure)
% \item science cases explored: spectroastrometry, imaging
% \item with 30-m class telescope, what can we achieve? 
% \end{itemize}

\section{SPECTROASTROMETRY}

Spectroastrometry is a technique of measuring the center of light as a function of wavelength \cite{bai98, whe04, whe08}. It is used for studying objects whose morphology changes with wavelength, on scales smaller than the resolution limit. A wavelength-dependent shift in the center of light --- the definition of spectroastrometric signal --- may indicate the presence of a companion, outflow, or any spatially extended feature. In \cite{kim22}, the use of PLs for spectroastrometry was studied. By measuring the wavelength-dependent intensities in output SMFs, the two-dimensional centroid shift can be recovered. 

At small phase aberration regime, the SMF output intensities of an $N$-port PL can be modeled as \cite{lin22a}
\begin{equation}
    \vec{I}(\vec{\phi}) \approx \vec{I}_0 + B \vec{\phi}
\end{equation}
where $\vec{I}(\phi)$ is the $N$ dimensional vector describing normalized output intensity (intensity in each port divided by the total intensity) with phase-only wavefront aberration of $\vec{\phi}$ ($M$ dimensional), $\vec{I}_0$ is the normalized output intensity of an on-axis point source, and $B$ is the $N\times M$ dimensional linear response matrix. For example, a 6-port PL ($N=6$) can sense the first five non-piston Zernike aberration modes ($M=5$). By measuring $\vec{I}(\vec{\phi}) - \vec{I}_0$ (normalized relative intensities), the phase aberration vector $\vec{\phi}$ can be recovered as
\begin{equation}\label{eq:recovery}
    \vec{\phi} \approx B^{+} (\vec{I}(\vec{\phi}) - \vec{I}_0)
\end{equation}
where $B^+$ denotes the pseudo-inverse of the $N \times M$ dimensional $B$ matrix.
The tip-tilt modes of $\vec{\phi}$ correspond to the centroid shift. By analyzing $\vec{I}(\vec{\phi}) - \vec{I}_0$ as a function of wavelength, spectroastrometric signals can be detected.

The spectroastrometric signal to noise, in the photon noise-limited regime, scales as \cite{whe08}:
\begin{equation}
    {\rm S/N} \sim \sqrt{N_{\rm phot}} \,\,\frac{{\rm center\,of\,light\,shift}}{\lambda/D} \propto D^2.
\end{equation}
Therefore, with 30\,m-class telescopes, an order of magnitude larger signal to noise is expected compared to that of current 10\,m-class telescopes. Ref. \cite{kim22} studied the possibility of detecting accreting planets using PL spectroastrometry of hydrogen emission lines. In this work, we demonstrate two other potential applications of PL spectroastrometry: detection of tidally heated exomoons and resolving broad-line regions of quasars. We use a numerically simulated 6-port PL transfer matrix model described in \cite{kim22} for the mock observations.

\subsection{Detection of exomoons}\label{ssec:exomoon}

% Kipping et al. 2022 \cite{kip22} surveyed 70 cool, giant transiting exoplanet candidates found by Kepler, and found one candidate Kepler 1708 b-i, transiting 2.6 Earth radii object orbiting a Jupiter sized host at 1.6 AU from the host. 

% \paragraph{Spectroastrometry}
% Agol et al. 2015 \cite{ago15} studied detectability of exomoons with spectroastrometry. 

The detectability of exomoons with spectroastrometry has been studied by \cite{ago15}, but to date, no exomoon candidates have been proposed using this technique. One potential target for direct light detection is tidally heated exomoons (THEMs) orbiting gas giants \cite{lim13}, with solar system analogs such as Io and Europa. They can possibly be more luminous than their host planet at some wavelengths (where the spectra of the host planet display absorption bands) in the mid-infrared and can shine brightly even at large separations from the stellar primary. Low-loss PLs operating in mid-infrared wavelengths can be enabled by ultrafast laser inscription techniques \cite{arr14}. For instance, \cite{kle23} studied observability of THEMs around $\epsilon$ Eridani b using JWST/MIRI. If the separation is favorable, disentangling the flux of the host planet and the exomoon may be possible using the spectroastrometric technique. Note that a Jupiter -- Io analog at the distance of Proxima Centauri (1.3\,pc) would have separation of about 2\,mas. 

One of the major advantages of using a PL for spectroastrometry is that two-dimensional high spectral resolution spectroastrometric signals can be detected simultaneously, while long-slit spectroastrometry measures spatial structures along one axis at a time and requires multiple position angles. Thus PL spectroastrometry may be more efficient for surveying exomoons around directly imaged planets. When the host planet light is coupled on-axis, the off-axis light from the exomoon will introduce wavelength-dependent variations in normalized intensities. 

\begin{figure} [hbt!]
\begin{center}
\begin{tabular}{c} 
\includegraphics[width=0.8\linewidth]{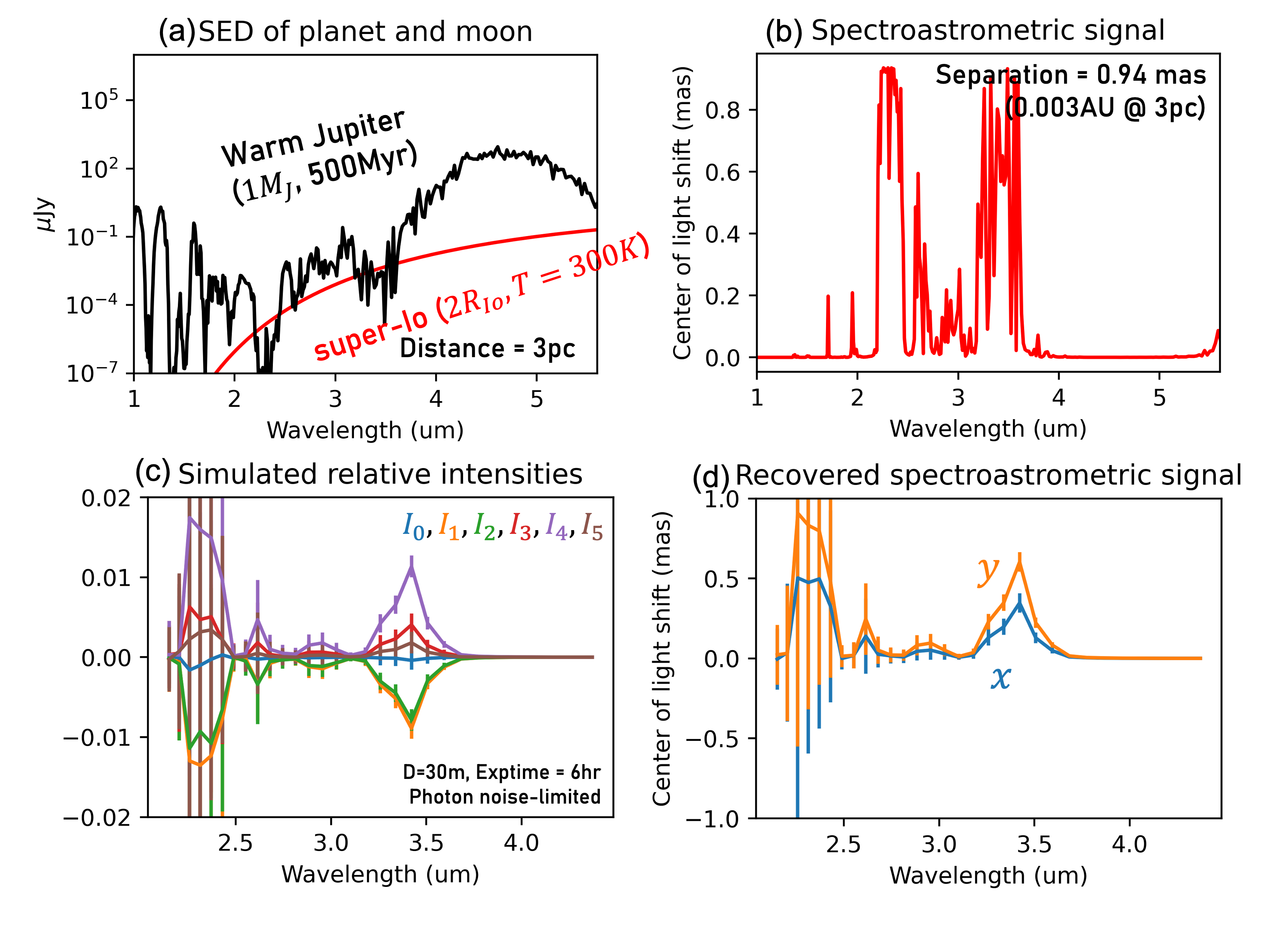}
\end{tabular}
\end{center}
\caption[example] 
{ \label{fig:exomoon} 
Mock observation of a super-Io around a warm Jupiter with separation of 0.94 mas and position angle of 60 degrees. (a) Spectral energy distribution of a warm Jupiter (1 Jupiter mass, 500 Myr) and a super-Io (2 Io radius, effective temperature of 300\,K) in near infrared. The super-Io outshines the host planet at methane and water absorption bands. (b) Simulated centroid shift as a function of wavelength, assuming a separation of 0.94 mas. (c) Simulated normalized relative intensities of the 6-port PL, $\vec{I}(\vec{\phi}) - \vec{I}_0$. (d) Recovered two-dimensional spectroastrometric signals (first two components of $B^{+} (\vec{I}(\vec{\phi}) - \vec{I}_0)$).}
\end{figure}

We perform a 6-hour mock observation of a super-Io (2$R_{\rm Io}$, $T=300K$) around a warm Jupiter at the distance of 3 pc (Figure \ref{fig:exomoon}) to illustrate this technique. 
Panel (a) shows the spectral energy distribution of the two sources. We use a 500 Myr old Jupiter-mass gas giant model from \cite{spi12} for the planet and a 2 Io radius blackbody model for the moon. Panel (b) shows the centroid shift as a function of wavelength. We assume the photon noise-limited case and do not account for wavefront errors nor the flux from the stellar primary. The light of the planet couples to the lantern at the center of the FMF entrance and the light from the exomoon couples off-axis, introducing variations in normalized relative intensities, as shown in panel (c). Finally in panel (d), the recovered spectroastrometric signals (tip-tilt modes of the phase aberrations) are displayed. High signal-to-noise centroid shifts at about 3.4 \textmu m are noticeable, the wavelength range where the super-Io outshines the host planet.

\subsection{Resolving BLR of quasars}\label{ssec:quasar}

The possibility of measuring the size of the broad-line region (BLR) of high-$z$ quasars and studying the kinematic structures using spectroastrometry has been studied by \cite{ste15}. The centroid of the BLR emission line varies as a function of velocity due to its extended structure. They prospect that the size of the BLRs of quasars out to $z\sim 6$ can be constrained with a 30\,m class telescope. 
Ref. \cite{bos21} presented a tentative detection (3.2$\sigma$) of spectroastrometric signals of quasar J2123-0050 ($z=2.3$) with Gemini/GNIRS and measured the size of the BLR as 454 \textmu as. 
With PLs, two-dimensional kinematic structures of BLRs may be constrained without necessitating multiple position angles.

\begin{figure} [hbt!]
\begin{center}
\begin{tabular}{c} 
\includegraphics[width=0.8\linewidth]{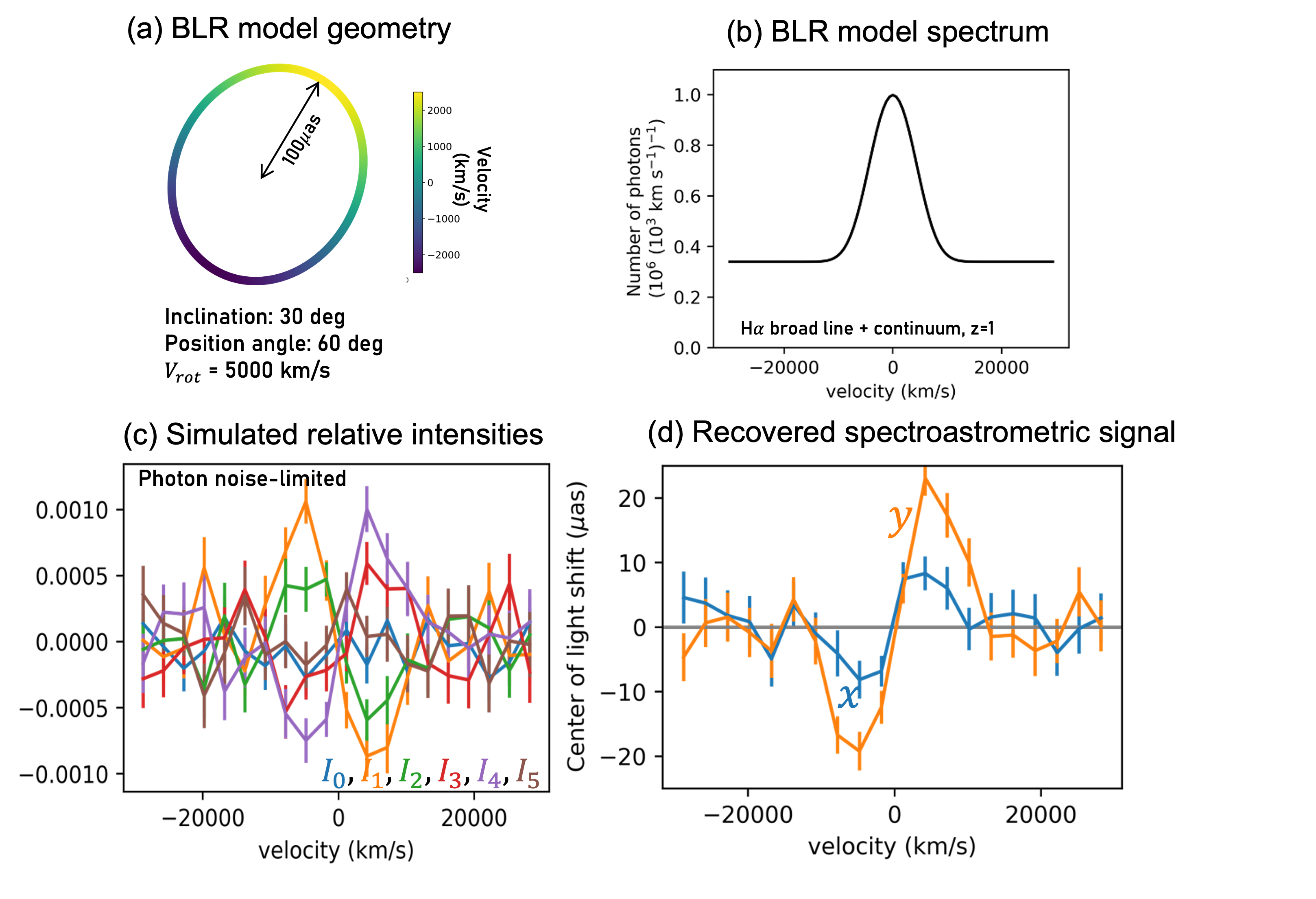}
\end{tabular}
\end{center}
\caption[example] 
{ \label{fig:blr} 
Mock observation of a BLR of a quasar. (a) The geometry of the simple ring BLR model. The color indicates the line of sight velocity. (b) The spectrum of the broad-line component of the H$\alpha$ line at near infrared ($z=1$) originating from the BLR. (c) Simulated normalized relative intensities of the 6-port PL, $\vec{I}(\vec{\phi}) - \vec{I}_0$. (d) Recovered two-dimensional spectroastrometric signals (first two components of $B^{+} (\vec{I}(\vec{\phi}) - \vec{I}_0)$).}
\end{figure}

Figure \ref{fig:blr} shows mock observation of a simplified BLR \cite{ste15}. The emission line originates from a thin ring of radius 100 \textmu as with the velocity dispersion of 2500 km s$^{-1}$ and rotational velocity of 5000 km s$^{-1}$ (panel (a)). The H$\alpha$ broad-line component spectrum is shown in the panel (b). Since the redshifted component and the blueshifted component of the emission line originate from different locations in the ring, the centroid of the emission line varies as a function of wavelength, which is reflected in the variation in normalized intensities in panel (c). Here we assume the photon noise-limited case and do not take wavefront error into account. The two-dimensional spectroastrometric signals can be recovered following Eq. \ref{eq:recovery} (panel (d)), which can then be used to infer the kinematic structure and position angle of the BLR. Although we considered a simple thin ring model in this work, the capability of measuring two-dimensional spectroastrometric signals will be particularly beneficial when fitting data to more complicated geometrical and dynamical models \cite{pan11, pan14}.

\section{Interferometric Imaging}\label{sec:imaging}

\begin{figure} [hbt!]
\begin{center}
\begin{tabular}{c} 
\includegraphics[width=0.8\linewidth]{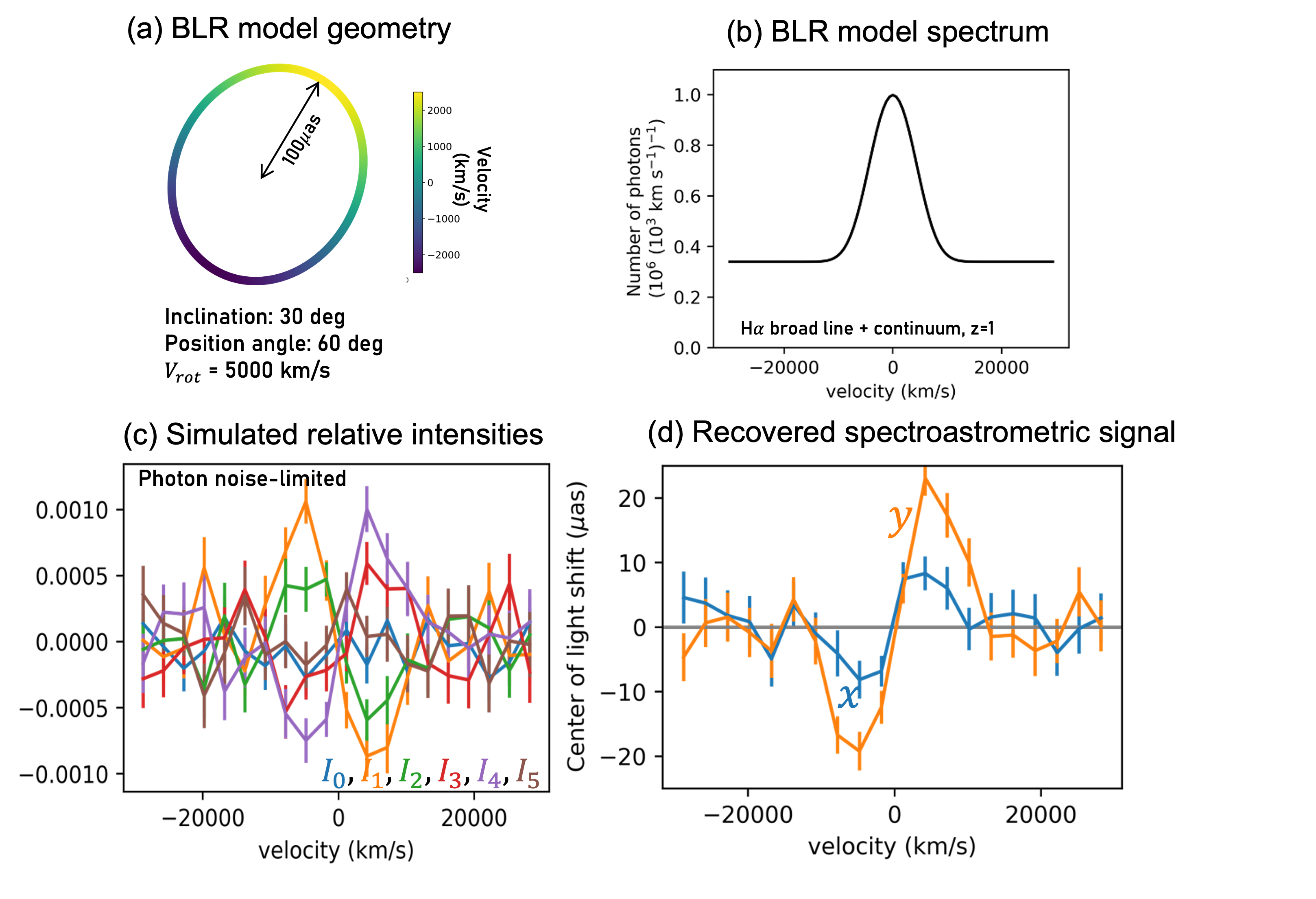}
\end{tabular}
\end{center}
\caption[example] 
{ \label{fig:interferometry} 
A conceptual diagram of a pairwise beam combiner for a 6-port PL, adopted from \cite{kim23spie}. Light from each single-mode waveguide is split five ways, and each of these is directed to ABCD beam combiners in a pairwise manner. The intensity of outputs from ABCD beam combiners can be used to determine the visibilities.}
\end{figure} 

\begin{figure} [hbt!]
\begin{center}
\begin{tabular}{c} 
\includegraphics[width=1.0\linewidth]{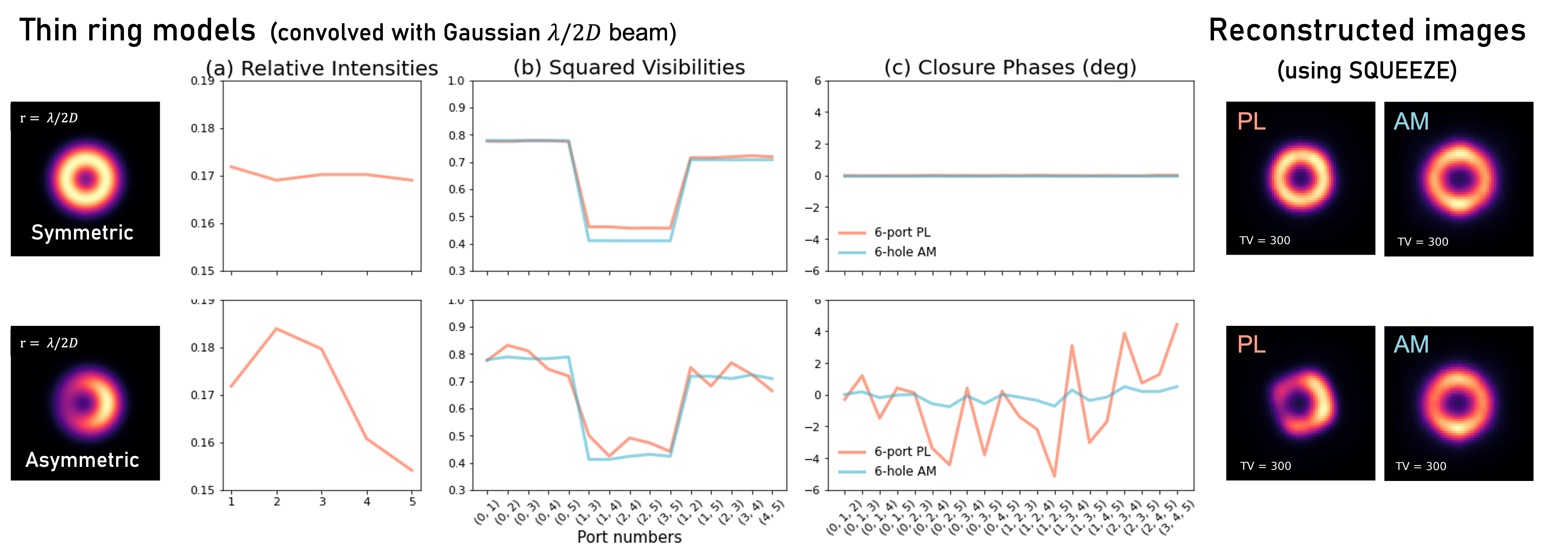}
\end{tabular}
\end{center}
\caption[example] 
{ \label{fig:ring} 
Simulated interferometric observables of a thin ring of $\lambda/2D$ radius with central unresolved point source. The upper panels and lower panels show the symmetric and asymmetric cases, respectively. The differences in signals between the symmetric and the asymmetric case are more significant for the PL than the conventional visibilities measured from an aperture masking interferometry analog (see \cite{kim23} for details). Therefore, reconstructed images using the modified {\tt squeeze} algorithm \cite{bar10} show that PLs can recover asymmetries better than the aperture masking interferometry analog.}
\end{figure}

By interfering SMF outputs of a PL, PLs can achieve higher sensitivity to smaller angular resolution scales compared to using only the SMF output intensities. Kim et al., submitted \cite{kim23} presented the concept of PL interferometry with a single telescope. In brief, the pupil plane PL principal modes (PLPMs), which are SMF fundamental modes backpropagated to the lantern entrance and then to the pupil plane, can be interpreted as the effective apertures of a PL. Therefore, interfering a pair of SMF outputs is analogous to interfering light from two different extended apertures (Figure \ref{fig:interferometry}). Ref. \cite{kim23} defined the PL visibilities as interferometric information measurable by pairwise beam combination of PL outputs. The PL visibilities are particularly sensitive to small angular scale asymmetries, as shown in simulated interferometric observables of thin symmetric and asymmetric rings with radius of $\lambda/2D$ (Figure \ref{fig:ring}). Here, $\lambda$ refers to wavelength and $D$ refers to the telescope diameter. Even at the resolution of $\lambda/2D$, the PLs can resolve asymmetric ring structures. With 30\,m-class telescopes, $\lambda/2D$ can reach a few mas in the near infrared. If the outputs are spectrally dispersed, spectro-interferometry can allow for even smaller angular resolutions \cite{des07}. This technique can image the very inner region of circumstellar disks as illustrated in Figure \ref{fig:ring}, and can search for close-in exoplanets.

\section{Summary and Future work}

This paper presents potential science applications of photonic lanterns (PLs) with extremely large telescopes. PLs can sense small angular scale variations in the input intensity distribution, suitable for observations that require resolving small angular scales. Two techniques are addressed: spectroastrometry and interferometric imaging. We showed photon noise-limited mock observations of tidally heated exomoons in section \ref{ssec:exomoon} and broad-line regions of quasars in section \ref{ssec:quasar}. In section \ref{sec:imaging} we briefly described the concept of interferometric imaging with PLs and the potential of imaging inner circumstellar disks. 

Our future work involves experimental validation of spectroastrometric techniques in laboratory and on-sky. Moreover, practical observing strategies and calibration methods in the presence of wavefront errors and misalignment need to be developed. Although we considered a standard 6-port PL in this paper, PL design parameters such as the number of cores and core sizes may be optimized \cite{lin23} for specific science applications.

% Practical complications include accounting for the variation in throughput of different PL outputs, dealing with chromaticity and polarization, and developing practical calibration methods in presence of wavefront errors and misalignment.

\acknowledgments % equivalent to \section*{ACKNOWLEDGMENTS}       
 
This work is supported by the National Science Foundation under Grant No. 2109231 and No. 2109232.
Y.J.K. thanks Hojin Cho for helpful discussions regarding BLR models.

% References

\printbibliography %Prints bibliography
\end{document}